\documentclass[twocolumn,pre,showpacs,,aps]{revtex4}
\usepackage{epsfig}

\begin{document}
\preprint{15-04-2009}

\title{Hysteresis in the $T=0$ RFIM: beyond metastable dynamics}
\author{Francesc Salvat-Pujol}
\affiliation{ Departament d'Estructura i Constituents de la Mat\`eria,
  Universitat de Barcelona \\ 
Mart\'{\i} i Franqu\`es 1, Facultat de F\'{\i}sica,
  08028 Barcelona, Catalonia}
\author{Martin-Luc Rosinberg}
\affiliation{Laboratoire  de  Physique  Th\'eorique  de  la  Mati\`ere
Condens\'ee,  CNRS-UMR 7600,  Universit\'e  Pierre et  Marie Curie,  4
place Jussieu, 75252 Paris Cedex 05, France}
\author{Eduard Vives}
%
%
\affiliation{ Departament d'Estructura i Constituents de la Mat\`eria,
  Universitat de Barcelona \\ 
Mart\'{\i} i Franqu\`es 1, Facultat de F\'{\i}sica,
  08028 Barcelona, Catalonia}
%

\begin{abstract}
We present a  numerical study of the zero-temperature  response of the
Gaussian random-field Ising model  (RFIM) to a slowly varying external
field, allowing the system to be trapped in microscopic configurations
that  are not  fully metastable.   This modification  of  the standard
single-spin-flip  dynamics  results  in  an  increase  of  dissipation
(hysteresis) somewhat  similar to that observed with  a finite driving
rate.   We  then  study  the  distribution  of  avalanches  along  the
hysteresis loop and perform  a finite-size scaling analysis that shows
good  evidence   that  the   critical  exponents  associated   to  the
disorder-induced phase transition are not modified.
\end{abstract}


\maketitle

\section{Introduction}

The  random-field Ising  model  (RFIM) at  zero  temperature has  been
proposed  as a  prototype  for  a broad  class  of disordered  systems
(random magnets, glasses,  plastic and ferro-elastic materials, fluids
in porous  media ...)  which exhibit an  intermittent, avalanche-like,
and    hysteretic   response   to    a   smoothly    varying   applied
force\cite{SDP2006}.   The   RFIM  has  also  been   used  in  several
socio-economics  contexts to  simulate collective  effects  induced by
imitation and  social pressure\cite{MB2005}.  A  remarkable prediction
of the model is the existence of a non-equilibrium critical point (for
a certain amount of disorder) which separates two different regimes of
avalanches.  In  the strong-disorder  regime,  all  avalanches are  of
microscopic    size    and     the    saturation    hysteresis    loop
(e.g. magnetization $m$ versus  magnetic field $H$) is macroscopically
smooth. At low  disorder, a macroscopic avalanche occurs  at a certain
field, which results in a  jump discontinuity in the magnetization. At
criticality,   avalanche   sizes   and  durations   follow   power-law
distributions.

In  the  original  version  of  the model\cite{S1993},  spins  obey  a
standard single-spin-flip  (Glauber) relaxation dynamics  at $T=0$ and
align with their local effective field. As the applied field is slowly
increased or decreased, a spin may become unstable and then trigger an
avalanche that propagates until another metastable state is found. The
field is  held fixed  during the propagation  of the  avalanche, which
corresponds  to the  so-called ``adiabatic''  limit.  This  amounts to
assuming that  the time  for local equilibration  (the duration  of an
avalanche)  is much smaller  than the  rate of  change of  the driving
field. Moreover,  by using a  deterministic zero-temperature dynamics,
one  assumes that  no thermally  activated escape  takes place  on the
observational time scale  (``athermal'' limit).  These two assumptions
are reasonable  in many  physical situations but  they are  never {\it
fully} satisfied.   It is thus  interesting to test the  robustness of
the   predicted   scenario  (in   particular   the   existence  of   a
disorder-induced phase  transition and its  universality) with respect
to a slight violation of these conditions.

In a recent work\cite{VRT2005}, the influence of partial equilibration
processes  was mimicked  by  changing the  dynamics  and allowing  two
neighboring  spins to  flip cooperatively.   As expected,  this change
resulted in  a reduction  of hysteresis (as  the set  of two-spin-flip
stable states is contained in  the set of one-spin-flip stable states)
and  an enhancement  of large-scale  collective effects\cite{IRT2006}.
But,  remarkably,  the  critical  behavior (characterized  by  various
exponents and  finite-size scaling functions)  remained identical.  In
the present  work, we  want to  go in the  opposite direction  by {\it
enlarging}  the  set of  visited  microscopic  states  so to  increase
hysteresis  and drive the  system further  from equilibrium.   This is
done  by allowing  a  small fraction  of  the spins  to  point in  the
opposite direction  to their  local field. However,  in order  to keep
things  as  simple  as  possible  (and, in  particular,  to  keep  the
simplifying  separation  of   motion  between  adiabatic  driving  and
avalanche propagation), we will  still use a single-spin-flip dynamics
and start an avalanche when some threshold in energy is reached.  As a
consequence, the saturation  loop will not be an  ``extremal'' path in
the field-magnetization plane\cite{PRRT2008,note1} and the property of
return-point  memory\cite{S1993} will be  violated. The  main question
that we want to address  is: will this change the universal properties
of  the  critical  behavior?   Note  that  this  modification  of  the
dynamical rule may  be viewed as a crude way  of simulating the effect
of a finite driving rate which does not give enough time to the system
to  relax, even  locally \cite{T1996,ZL2002,WD2003,PTMPV2004,CDZ2006}.
But this is  of course a caricature of what happens  in the real world
and we  therefore do not pretend  to propose here a  theory of dynamic
hysteresis,  a  topic  that   has  been  (and  still  is)  intensively
investigated  in  the  literature\cite{CA1999}.  In  a  socio-economic
context, one could  also interpret this model as  simulating an effect
of  ``inertia'' that  prevents  the individual  agents  to make  their
decision (e.g., buy or sell)  when the incentive reaches the threshold
value.

The  paper is  organized  as  follows.  In  Section  \ref{Model} ,  we
describe  the model  and the  new dynamics,  and discuss  some  of its
properties.  In  Section  \ref{Hysteresis},  we study  the  hysteresis
loops, in particular the change in coercivity. Avalanche distributions
are analyzed in Section  \ref{Avalanches} and the critical behavior in
Section \ref{Critical}.   Summary and conclusion are  given in Section
\ref{Conclusion}.

\section{Model and dynamics}
\label{Model}
We consider  the RFIM on  a three-dimensional cubic lattice  of linear
size $L$ with periodic boundary conditions.  On each site $i$ there is
a spin variable $S_i=\pm 1$. The energy of the system of $N=L^3$ spins
is described by the Hamiltonian
\begin{equation}
\label{Eq1}
{\cal H} = -J\sum_{<ij>}S_i S_j - \sum_{i} h_i S_i -H \sum_{i} S_i,
\end{equation}
where $J>0$  is a ferromagnetic exchange interaction  constant and the
first  sum   runs  over  nearest-neighbor  pairs.    Without  loss  of
generality, we will take $J=1$.  The set of random fields $\{h_i\}$ is
drawn independently from a Gaussian probability distribution with zero
mean and variance  $\sigma^2$, and $H$ is a  uniform external magnetic
field  which couples to  the overall  magnetization $M=  \sum_{i} S_i$
($m=M/N$).

The energy change associated to the reversal of spin $i$ then reads
\begin{equation}
\label{Eq2}
\Delta{\cal H}_i  \equiv \Delta{\cal H} (S_i \rightarrow  -S_i)= 2 S_i
F_i,
\end{equation}
where
\begin{equation}
\label{Eq3}
F_i = \sum_{j/i}S_j + h_i + H
\end{equation}
is  the  local effective  field  acting  on spin  $i$,  and  $j$ is  a
nearest-neighbor of $i$.

At $T=0$ the standard single-spin-flip dynamics consists in flipping a
spin   if  this   lowers  its   energy.   A   configuration   is  thus
single-spin-flip stable if every spin is aligned with its local field,
{\it i.e.},
\begin{equation}
\label{Eq4}
S_i = \text{sign} \left ( F_i \right ) \quad \forall i .
\end{equation}
Each  of these  metastable states  has  a certain  range of  stability
$H_{min}\le H\le H_{max}$.

As the  field $H$ is slowly  changed, a spin flips  (either upwards or
downwards) when its local field changes sign, which may induce a whole
avalanche of  other spin  flips.  In the  adiabatic limit $H$  is held
constant during the propagation  of the avalanche. The avalanche stops
when  a new  metastable  state is  reached.   A nice  feature of  this
dynamics  is  that  it   is  ``abelian'':  because  of  the  so-called
``no-passing'' rule\cite{S1993},  the metastable state  reached at the
end  of an avalanche  is independent  of the  order in  which unstable
spins have been  reversed. In consequence, these spins  can be flipped
either  sequentially or  in parallel  (this latter  choice  having the
advantage that one can define the duration of an avalanche).

We  now modify this  dynamical rule  by allowing  a certain  number of
spins to be unstable (Eq.~(\ref{Eq4})  is then violated).  This can be
done in  various ways, for instance  by imposing that  the fraction of
unstable spins  cannot exceed a  certain value.  This,  however, would
prevent the system to reach saturation. We therefore prefer to compute
the extra amount of energy associated to the unstable spins and impose
that  an avalanche starts  when this  contribution exceeds  some fixed
threshold  $\epsilon$.   Specifically,   the  system  may  visit  spin
configurations that we call ``$\epsilon$-stable'' and that satisfy
\begin{equation}
\label{Eq5}
\sum_{S_i \text{unstable}}\Delta {\cal H}_i \ge \ - N \epsilon \ ,
\end{equation}
where  the sum runs  over all  unstable spins,  i.e.  spins  for which
$\Delta{\cal H}_i<0$ (note that  $\epsilon$ is an intensive quantity).
There are now two conditions for an avalanche to start: (i) there must
be  unstable spins  (ii) the  sum of  the extra  contributions  to the
energy  due  to the  unstable  spins  must  exceed the  threshold  (in
absolute value).  When increasing  (resp.\  decreasing) the field, only
negative (resp.\   positive) spins contribute  to this energy.   One of
course recovers the usual dynamics for $\epsilon=0$.
\begin{figure}[hbt]
\begin{center}
\epsfig{file=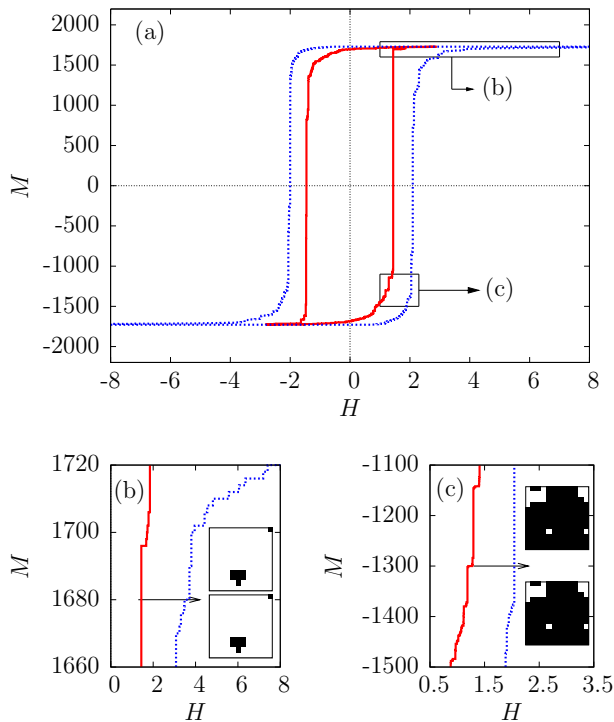,width=8.0cm,clip=}
\end{center}
\caption{\label{FIG1}(Color online) Comparison of the hysteresis loops
obtained  with  $\epsilon=0$  (solid  line)  and  $\epsilon=0.03$
(dashed line) in a system  of linear size $L=12$.  In (b) and (c)
the  upper and lower  parts of  the ascending  branches are  blown up,
showing  that  some  avalanches  merge  or split  when  $\epsilon$  is
changed.   For   a  given  overall  magnetization   $M$,  the  reached
configurations are identical (as illustrated by two-dimensional slices
of  the system  where  negative spins  are  drawn in  black), but  the
corresponding fields are different.}
\end{figure}

It is quite obvious that  this modification of the dynamics spoils the
no-passing rule  and the  abelian property.  Therefore,  since several
spins may be  unstable when the energy threshold  is reached, one needs
to specify  the order in which  these spins are  flipped.  The natural
choice that we adopt is to  flip the most unstable spin first, that is
the one that  corresponds to the most negative  $\Delta{\cal H}_i$. We
then update the local fields  while keeping $H$ constant, search again
for  the most  unstable spin,  flip it,  etc., until  Eq.~\ref{Eq5} is
satisfied\cite{note0}.  This  ``greedy'' algorithm essentially amounts
to  performing a  steepest descent  path  in energy,  which defines  a
deterministic sequence of unstable states inside an avalanche (defined
as usual as the collection of spins which flip at the same field).  It
is then  not difficult to  see that the  order in which the  spins are
flipped along the hysteresis loop {\it does not} depend on $\epsilon$,
so  that  the  sequence  of  visited states  is  invariant.   Changing
$\epsilon$ only changes the values  of the external field at which the
states are reached  or left.  More generally, in  any monotonous field
history  from  a  state  with   magnetization  $M$  to  a  state  with
magnetization $M'$,  the sequence of visited  spin configurations does
not depend on $\epsilon$, but the corresponding fields may differ.  As
a result,  avalanches may split  or merge when changing  $\epsilon$ so
that  their number  and  size  also change.   This  is illustrated  in
Fig.~\ref{FIG1}  where we  compare the  hysteresis loops  obtained for
$\epsilon=0$ and $\epsilon=0.03$ in a  small size system.  One can see
in Figs.~\ref{FIG1}(b) and (c) that the spin configurations at a given
magnetization $M$  are identical although  the hysteretic trajectories
are different.   Fig.~\ref{FIG1}(b) also shows  that a state  which is
not single-spin-flip stable for $\epsilon=0$  (as it is located in the
middle of an  avalanche) is ``$\epsilon$-stable'' for $\epsilon=0.03$.
In Fig.~\ref{FIG1}(c), the opposite situation is observed.

\begin{figure}[hbt]
\begin{center}
\epsfig{file=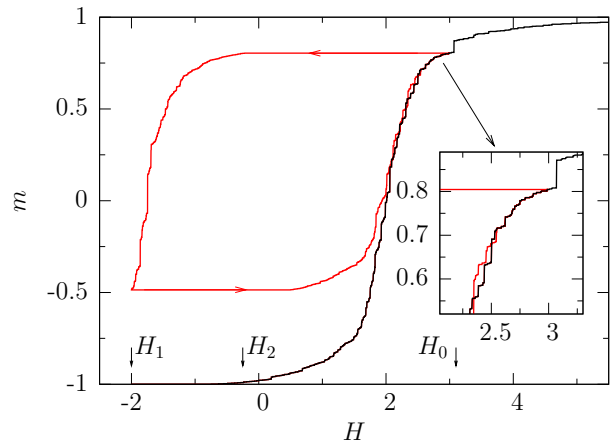, width=8cm,clip=}
\end{center}
\caption{\label{FIG2}(Color  online) Test  of the  return-point memory
property  in a  system  of  linear size  $L=12$  for $\sigma=3.5$  and
$\epsilon=0.05$. The  field is increased  up to $H_0$,  decreased from
$H_0$ to $H_1$ and then  increased again up to saturation.  The return
trajectory crosses  the ascending branch of  the saturation hysteresis
loop  several times, indicating  that $\epsilon$-states  exist outside
the loop and that the RPM property is violated.}
\end{figure}

The fact  that the modified  dynamics does not satisfy  the no-passing
rule has two consequences.  Firstly, there can exist $\epsilon$-states
outside  the hysteresis loop\cite{note1}.   Secondly, the  property of
return-point   memory   (RPM)   is   not  satisfied,   as   shown   in
Fig.~\ref{FIG2}. However, the violation of RPM is small for the values
of $\epsilon$ considered in this  work and it seems that this property
is better  and better verified as  the system size  increases. One can
also see in Fig.~\ref{FIG2}  that the first-order reversal curves have
a linear portion: when reversing the field (for instance from $H_0$ to
$H_1$), the negative  spins that were unstable at  $H_0$ become stable
again before  any positive spin becomes unstable  and Eq.~\ref{Eq5} is
only violated when $H<H_2$.

As a  final remark in this  Section, we compare  the proposed dynamics
with the  algorithms that have been  used previously to  study the T=0
RFIM with  a finite driving rate\cite{T1996,WD2003,PTMPV2004,CDZ2006}.
The most  common strategy consists  in increasing the field  in finite
steps $\Delta H$, thus merging all the avalanches occurring within that
field  window into  a  larger avalanche\cite{T1996,PTMPV2004,CDZ2006}.
The resulting magnetization loops then share a series of common points
with those obtained with the  adiabatic driving. On the other hand, in
our case  avalanches not only merge  but also split  and the resulting
loops differs  everywhere from the adiabatic  ($\epsilon=0$) loops.  A
second strategy\cite{WD2003,PTMPV2004} consists in performing an exact
simulation of the  continuous $M(t)$ signal by using  a finite driving
rate and  defining a  time interval associated  to the shell  of spins
that  relax in  parallel. But  one then  needs to  fix a  threshold to
define  the  avalanches and  this  has  a  strong influence  on  their
size\cite{PBB2005}.   This  problem does  not  occur  in the  modified
dynamics that we used here  since there is still a complete time-scale
separation  between  the field  driving  and  the  propagation of  the
avalanches.

\section{Hysteresis loops}
\label{Hysteresis}

\begin{figure}[hbt]
\begin{center}
\epsfig{file=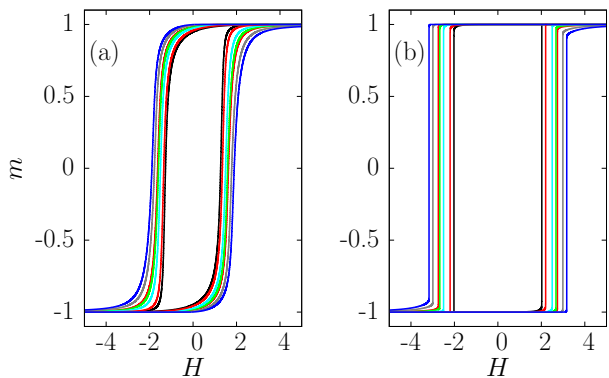, width=8cm,clip=}
\end{center}
\caption{\label{FIG3}  (Color  online)  Average  hysteresis  loop  for
$\sigma=2.8$  (a)   and  $\sigma=1.5$  (b).    The  curves  correspond
respectively   to   $\epsilon=0$   (inner   loop),   $0.0005,   0.004,
0.008,0.01,0.02$, and $0.03$ (outer loop). Data result from an average
over $5000$ disorder realizations in a system of size $L=24$. }
\end{figure}

We  first consider the  influence of  $\epsilon$ on  the shape  of the
hysteresis  loops  for  different  values  of  the  disorder  strength
$\sigma$. Fig.~\ref{FIG3} shows the results obtained by averaging over
different  disorder realizations for  $\sigma=3$ and  $\sigma=1.5$. As
expected, the main effect of $\epsilon$ is to bring the system further
away from equilibrium and  to increase hysteresis.  For $\sigma=3$ for
instance, the  loop area, which represents the  energy loss, increases
by $\sim70\%$ when increasing $\epsilon$  from $0$ to $0.03$.  This is
already an important variation and  in the following we shall restrict
our study to the range $0\le \epsilon \le 0.03$.

Fig.~\ref{FIG3} also shows that  there are still two different regimes
when  $\epsilon\ne 0$ and  that the  shape of  the loops  changes from
smooth to  rectangular as $\sigma$  decreases.  In particular,  at low
disorder, there is a single avalanche that spans the whole system at a
certain value of the external  field (note that the spanning avalanche
in   a  finite  system   occurs  with   no  collective   precursor  for
$\epsilon>0.03$, which may change its nature. This is also a reason to
restrict our study to smaller values of $\epsilon$).

\begin{figure}[ht]
\begin{center}
\epsfig{file=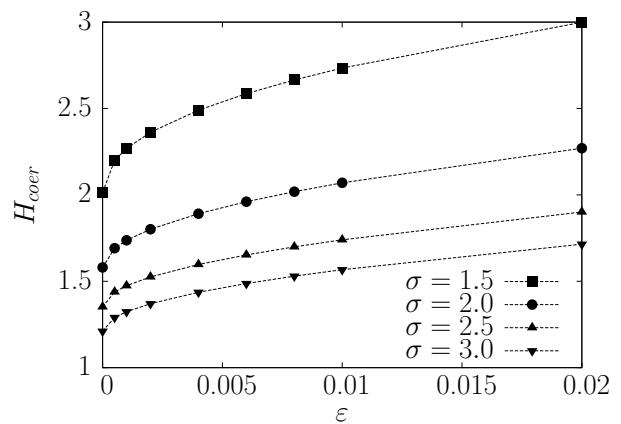, width=8cm,clip=}
\end{center}
\caption{\label{FIG4}  Average   coercive  field  as   a  function  of
$\epsilon$ for different  values of $\sigma$. The data  result from an
average over  $500$ disorder realizations  in a system of  size $L=24$
and are fitted according to Eq.~(\ref{Eq10}).}
\end{figure}

To further quantify the influence  of $\epsilon$ on the hysteresis, we
show in Fig.~\ref{FIG4} the variation of the coercive field $H_{coer}$
for different values of the disorder.  We find that the data are quite
accurately fitted by the equation
 \begin{equation}
\label{Eq10}
H_{coer}(\epsilon)= H_{coer}(0)+C \epsilon^{\beta},
\end{equation}
with  $\beta \approx  0.5$  in the  large-disorder  regime and  $\beta
\approx  0.45$ at  low disorder\cite{note2}.   The same  dependence is
found for the variations of the loop area with $\epsilon$.  So far, we
have no convincing theoretical  explanation for this behavior.  On the
other hand, it appears that the same exponent $\beta \approx 0.45$
in the low disorder regime has  been observed in a simulation study of
the  RFIM under  a linear  driving rate\cite{ZL2002}  (in  that study,
however,  there  is no  clear  scaling  at  large disorder).   In  two
dimensions\cite{CDZ2006}, when varying the  field in small steps as in
Ref.~\onlinecite{PTMPV2004},  simulations  show  a  crossover  from  a
square-root to a linear dependence  of $H_{coer}$ with the rate as the
disorder  is increased,  in agreement  with the  behavior  observed in
ferromagnetic thin films.

\section{Avalanche size distribution}
\label{Avalanches} 

The  RFIM  with  the   standard  ($\epsilon=0$)  dynamics  displays  a
power-law  distribution  of avalanche  sizes  at  a critical  disorder
$\sigma=\sigma_c$\cite{S1993}.   Depending  on   the  method  used  to
extrapolate  the numerical  results  to the  thermodynamic limit,  the
value  of  $\sigma_c$  varies  from $2.16$  \cite{PDS1999}  to  $2.21$
\cite{PV2003,PV2004}.  In  this section we  study the behavior  of the
avalanche size distribution for $\epsilon>0$.
\begin{figure}[ht]
\begin{center}
\epsfig{file=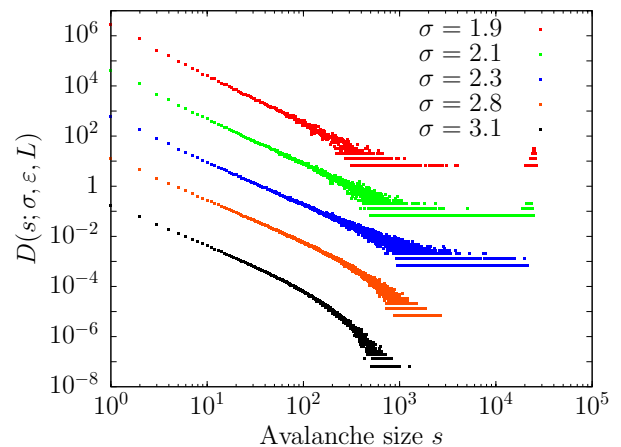, width=8cm,clip=}
\end{center}
\caption{\label{FIG5} (Color online) Avalanche size distributions in a system
of size $L=30$  for $\epsilon=0.008$  and different  values of $\sigma$.  The
curves are sorted from top to bottom in the order indicated in the legend. The statistics  has
been performed  over $1500$  disorder realizations.   Data   for  $\sigma
\neq 2.3$  have   been  shifted vertically.} 
\end{figure}

\begin{figure}[ht]
\begin{center}
\epsfig{file=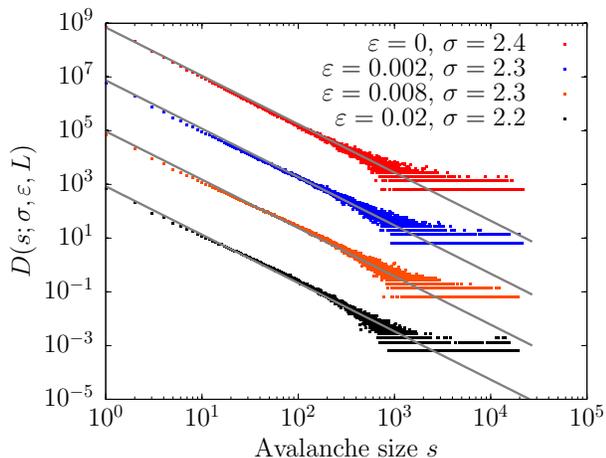, width=8cm,clip=}
\end{center}
\caption{\label{FIG6} (Color online) Avalanche size distributions in a
system of  size $L=30$  for different values  of $\epsilon$. The curves are
sorted from top to bottom in the order indicated in the legend. In each
case, the disorder  $\sigma$ is the one for  which the distribution is
closest to a  power-law. The slope of the  dashed lines that describe
the power-law region is $-1.8$. The statistics has been performed over
$1500$ disorder realizations. Data  for $\epsilon \neq 0.02$ have been
shifted vertically.}
\end{figure}

Fig.~\ref{FIG5}     shows    the    avalanche     size    distribution
$D(s;\sigma,\epsilon,L)$  obtained  in a  system  of  size $L=30$  for
$\epsilon=0.008$ and various  disorders $\sigma$ when sweeping through
a half-loop.  The  same behavior as for $\epsilon=0$  is observed: for
large $\sigma$  the distribution  is exponentially damped  whereas for
small $\sigma$ there  is a peak at large sizes  due to avalanches with
characteristic  size  $\sim   L^3$.   These  spanning  avalanches  are
responsible for the macroscopic  discontinuity in the hysteresis loop.
Between these two regimes, there is  a value of $\sigma$ for which the
distribution  is very  well  approximated  by a  power-law  up to  the
trivial cutoff size $s_{max}\sim L^3$ (and apart from some corrections
at very small $s$).  This  ``critical'' value of $\sigma$ changes with
$\epsilon$  but  the slope  of  the  power-law  region appears  to  be
invariant, as  shown in  Fig.~\ref{FIG6},.  We view  this result  as a
first   indication    that   there   exists    a   critical   disorder
$\sigma_c(\epsilon)$ in the thermodynamic limit and that the power-law
exponent  $\tau+\sigma\beta\delta$  that  characterizes the  avalanche
size  distribution  at criticality\cite{S1993}  does  not change  with
$\epsilon$ (note that the numerical  value obtained in a finite system
may differ from the actual  value $\sim 2$ in the thermodynamic limit,
as  estimated in  Ref.~\onlinecite{S1993}).   In the  next Section  we
present  a finite-size  scaling  analysis that  will corroborate  this
statement.

Note that this result contrasts with the one that has been obtained in
previous     studies      using     a     finite      driving     rate
\cite{WD2003,PTMPV2004}. Indeed,  when the  only effect of  the finite
driving  rate is  to  merge  small avalanches  into  larger ones,  the
power-law  exponent  decreases.  It  is  in  fact  unclear if  such  a
decrease  is  due to  the  actual  out-of-equilibrium  behavior or  is
induced  by  the  approximate   treatment  of  the  avalanche  merging
phenomenon and/or  by the definition  of the threshold that  allows to
discriminate the avalanches.

\section{Critical properties}
\label{Critical}

The analysis of the spanning  avalanches has proven to be a successful
way to determine the values  of several critical exponents in the RFIM
with  the standard  metastable  dynamics\cite{PV2003,PV2004}.  Indeed,
the  statistics  of spanning  avalanches  in  finite systems  contains
information   about  the   percolating  fractal   avalanches   in  the
thermodynamic limit  which are the signature of  criticality.  In this
Section we  use a  finite-size scaling method  to study the  effect of
$\epsilon$ on the critical behavior of the model.
\begin{figure}[ht]
\begin{center}
\epsfig{file=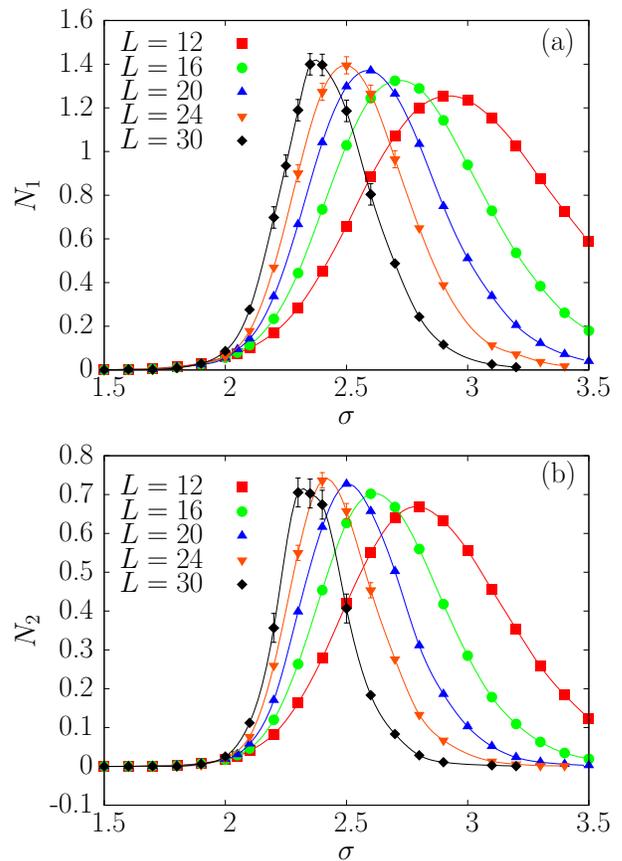, width=8cm,clip=}
\end{center}
\caption{\label{FIG7} (Color online) Average number of 1D- (a) and 2D-
(b) spanning avalanches as a function of $\sigma$ for $\epsilon=0.008$
and different system  sizes. Typical error bars for  the largest sizes
are shown.}
\end{figure}

With the modified dynamics  ($\epsilon>0$) an avalanche still involves
a connected set of spins and the spanning avalanches can be defined as
usual  : avalanches  that span  the  whole system  in $1,  2$, or  $3$
spatial  dimensions  are referred  to  as  1D-,  2D-, and  3D-spanning
avalanches, respectively. We focus  our analysis on the average number
of 1D- and 2D-spanning avalanches  occurring along the lower branch of
the  hysteresis  loop,  which  we  call  $N_1(\sigma,\epsilon,L)$  and
$N_2(\sigma,\epsilon,L)$, respectively.  We  discard from our analysis
the 3D-spanning  avalanches, which contain information  not only about
the  critical  percolating  avalanches,  but also  about  the  compact
infinite avalanche that gives rise to the first-order discontinuity in
the low-disorder regime\cite{PV2003}.
\begin{figure}[ht]
\begin{center}
\epsfig{file=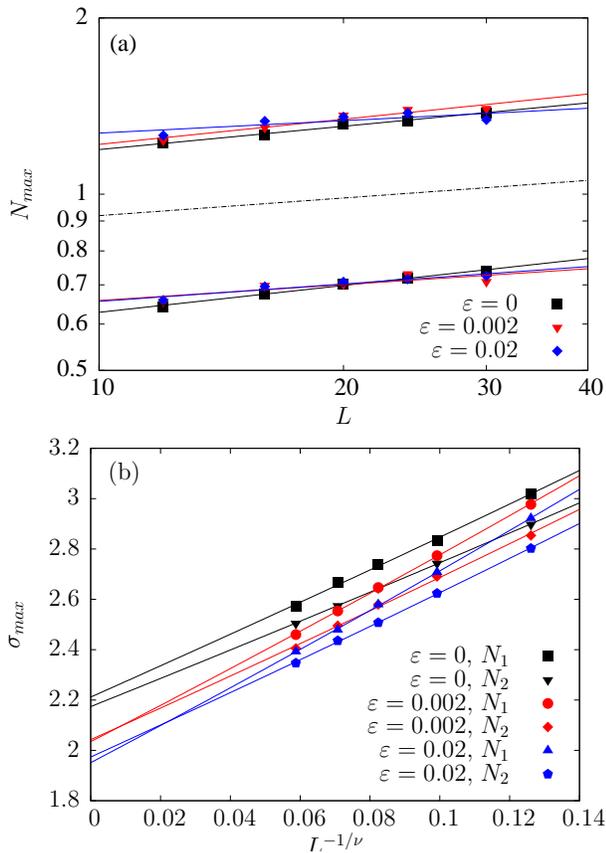, width=8cm,clip=}
\end{center}
\caption{\label{FIG8} (Color  online) Height and position  of the peak
in $N_1(\sigma,\epsilon,L)$ and $N_2(\sigma,\epsilon,L)$ for different
values of $\epsilon$  and different system sizes.  In  (a), the height
is  plotted  vs.   $L$  in a log-log plot and   in  (b)  the  position  is  plotted  vs.
$L^{-1/\nu}$  with  $\nu=1.2$.   The  data  are  fitted  according  to
Eqs.(\ref{scal1}) with $A=-0.1$. The dotted line in (a) has a slope 0.1 and is
included as a reference.}
\end{figure}

Figure \ref{FIG7} illustrates the behavior of $N_1(\sigma,\epsilon,L)$
and   $N_2(\sigma,\epsilon,L)$   as  a   function   of  $\sigma$   for
$\epsilon=0.008$ and  different system  sizes (the data  correspond to
averages over  typically $10^3-10^4$ disorder  realizations).  One can
see that  both functions  exhibit a peak  whose height  increases with
$L$. Moreover the  peak position shifts and its  width reduces.  These
features are clear  signatures of the existence of  an infinite number
of   percolating   avalanches   at   a   certain   critical   disorder
$\sigma_c(\epsilon)$       in      the       thermodynamic      limit.
$N_1(\sigma,\epsilon,L)$   and   $N_2(\sigma,\epsilon,L)$   are   thus
expected to have the scaling form\cite{PDS1999,PV2003}
\begin{eqnarray}
\label{scal1}
N_1(\sigma,\epsilon,L)&     =      &     L^{\theta}     {\hat     N}_1
(uL^{1/\nu},\epsilon), \\
\label{scal2}
N_2(\sigma,\epsilon,L)&     =      &     L^{\theta}     {\hat     N}_2
(uL^{1/\nu},\epsilon) \ ,
\end{eqnarray}
where  $\hat N_1$ and  $\hat N_2$  are finite-size  scaling functions,
$u(\sigma)$  is  some  analytical  function  of the  distance  to  the
critical disorder, and $\theta$  and $\nu$ are critical exponents that
characterize the  divergence of the number of  spanning avalanches and
of  the  correlation  length,  respectively.   Although  the  function
$u(\sigma)$     can    be     approximated    to     first-order    as
$(\sigma-\sigma_c(\epsilon))/\sigma_c(\epsilon)$,   previous   studies
\cite{PV2003} suggest that higher order terms are necessary to produce
good scaling collapses.  Therefore, as in Ref.~\onlinecite{PV2003}, we
shall use
\begin{equation}
u=\frac{\sigma-\sigma_c(\epsilon)}{\sigma_c(\epsilon)}+     A(\epsilon)
\left  (  \frac{\sigma-\sigma_c(\epsilon)}{\sigma_c(\epsilon)}  \right
)^2,
\end{equation}
where  $A(\epsilon)$ is a  nonuniversal parameter  that may  depend on
$\epsilon$.   For  $\epsilon=0$,  the  best  choice was  found  to  be
$A=-0.2$. This  value did not change when  replacing the $1$-spin-flip
by the $2$-spin-flip dynamics\cite{VRT2005}.

Equations  (\ref{scal1})  and (\ref{scal2})  may  be  first tested  by
plotting  the  height of  the  peaks  in $N_1(\sigma,\epsilon,L)$  and
$N_2(\sigma,\epsilon,L)$ as a  function of $L$ in a  log-log scale. As
shown in Fig.~\ref{FIG8}(a), the behavior is then linear and the slope
is   compatible    with   the   value    $\theta=0.1$   obtained   for
$\epsilon=0$\cite{PV2003}.

A second test consists in  plotting the position $\sigma_{max}$ of the
peaks as  a function of $L^{-1/\nu}$, as  shown in Fig.~\ref{FIG8}(b).
From  Eqs.~(\ref{scal1}) and  (\ref{scal2}), this  position  should be
determined by the condition $u L^{-1/\nu} =$ constant, {\it i.e.},
\begin{equation}
\label{scalK}
KL^{1/\nu}=\frac{\sigma_{max}(\epsilon)-\sigma_c(\epsilon)}
{\sigma_c(\epsilon)}+    A   \left    (   \frac{\sigma_{max}(\epsilon)
-\sigma_c(\epsilon)}{\sigma_c(\epsilon)} \right )^2.
\end{equation}
For $A=0$  this equation predicts a linear  behavior of $\sigma_{max}$
as a function  of $L^{-1/\nu}$. As can be  seen in Fig.~\ref{FIG8}(b),
the  actual behavior  is indeed  almost  linear when  using the  value
$\nu=1.2$ obtained for $\epsilon=0$\cite{PV2003}  and the data for the
1D- and 2D-spanning avalanches reasonably extrapolate towards the same
value in  the thermodynamic limit $L\rightarrow  \infty$. This method,
however,  cannot  be  used  to  extract  accurate  values  of  $A$  or
$\sigma_c(\epsilon)$  and  the fits  shown  in Fig.~\ref{FIG8}(b)  are
based on the  results of the finite-size scaling  analysis that we now
discuss.
\begin{figure}
\begin{center}
\epsfig{file=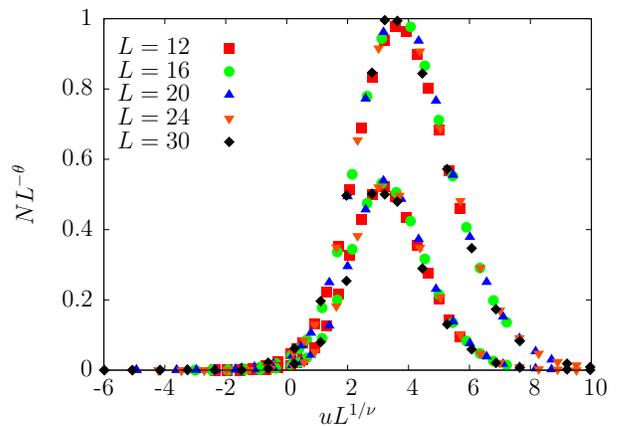, width=8cm,clip=}
\end{center}
\caption{\label{FIG9} (Color online) Scaling plot of the number of 1D-
and 2D-spanning  avalanches for $\epsilon=0.008$  and different system
sizes.  The upper  (resp.\  lower) curve corresponds to  the 1D- (resp.
2D-)    spanning   avalanches    using    $\sigma_c=1.97$,   $A=-0.1$,
$\theta=0.1$, and $\nu=1.2$.}
\end{figure}

Indeed,  the  best way  to  estimate  all  universal and  nonuniversal
parameters  is   to  search  for   a  good  collapse  of   the  curves
$N_1(\sigma,\epsilon,L)$ and $N_2(\sigma,\epsilon,L)$ corresponding to
different sizes $L$.  This is done by plotting $N_1(\sigma,\epsilon,L)
L^{-\theta}$ as a function of  the scaling variable $u L^{1/\nu}$.  As
an example, we show in  Fig.~\ref{FIG9} the best collapse obtained for
$\epsilon=0.008$   using   the   values   $\sigma_c=1.97$,   $A=-0.1$,
$\theta=0.1$, and $\nu=1.2$.  This procedure can be done independently
for  each set of  data corresponding  to different  $\epsilon$.  Table
\ref{Best} shows the parameters that produce the best collapses of the
curves for $0.0005\le \epsilon \le 0.03$\cite{note5} For comparison we
also include  the results obtained  for $\epsilon=0$\cite{PV2003}.  In
all cases,  the values $\nu=1.2$,  $\theta=0.1$, and $A=-0.1$  are the
best  estimates. The only  parameter showing  a clear  dependence with
$\epsilon$ is  $\sigma_c$.  Notice that  $A$, which in principle  is a
nonuniversal  parameter,   takes  the   same  value  $-0.1$   for  all
$\epsilon>0$. For  $\epsilon=0$ the  value $A=-0.2$ produced  a better
collapse  \cite{PV2003} but  the  difference is  not very  significant
since the collapse in Fig.~\ref{FIG9}  is still rather good when using
this   value   (alternatively,  one   can   also   use  $A=-0.1$   for
$\epsilon=0$).

\begin{table}
\begin{tabular}{r|r|r|r|r|r}
$\epsilon$ & $\nu$ &  $\theta$ & $\sigma_c$ & $A$ & $B$  \\ \hline 0 &
1.2 & 0.1 & 2.21  & -0.2 & 1.26 \\ 0.0005 & 1.2 &  0.1& 2.035 & -0.1 &
1.05 \\ 0.001 & 1.2 & 0.1 & 2.02  & -0.1 & 1.04 \\ 0.002 & 1.2 & 0.1 &
2.02 & -0.1 & 1.02 \\ 0.006 & 1.2  & 0.1 & 1.98 & -0.1 & 1.01 \\ 0.008
& 1.2 & 0.1 &  1.97 & -0.1 & 1.00 \\ 0.01 & 1.2 &  0.1 & 1.97 & -0.1 &
1.01 \\ 0.02  & 1.2 & 0.1 & 1.96 &  -0.1 & 1.02 \\ 0.03 &  1.2 & 0.1 &
1.97 & -0.1 & 1.02
\end{tabular}
\caption{\label{Best} Universal and nonuniversal parameters that yield
the  best  finite-size  scaling  collapses  for  different  values  of
$\epsilon$.}
\end{table}

\begin{figure}
\begin{center}
\epsfig{file=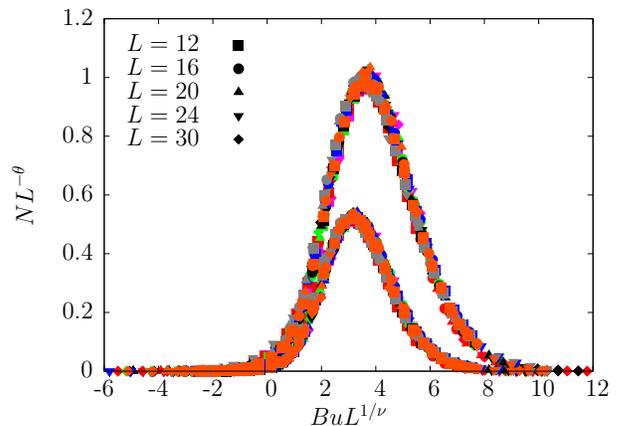, width=8cm,clip=}
\end{center}
\caption{\label{FIG10} (Color  online) Scaling  plot of the  number of
1D-  (upper  curves)  and   2D-  (lower  curves)  spanning  avalanches
according  to  Eqs.~(\ref{scal21})  and (\ref{scal22})  for  different
system sizes $L$ (indicated by different symbols) and for
$\varepsilon=0.001,0.002,0.006,0.008,0.01,0.02,0.03$ (not indicated).}
\end{figure}

The  most  important feature  in  Table  \ref{Best}  is that  the  two
critical exponents  $\nu$ and $\theta$  do not change  with $\epsilon$
and are the same as  for $\epsilon=0$. This suggests that the critical
behavior is described by the  same universality class.  We then try to
collapse all  the data  for different $\epsilon$  and $L$ on  the same
plot  by   introducing  a  nonuniversal  scale  factor   $B$  that  is
$\epsilon$-dependent, {\it i.e.}, by assuming that
\begin{eqnarray}
\label{scal21}
N_1(\sigma,\epsilon,L)&  =   &  L^{\theta}  {\hat   N}_1  (B(\epsilon)
uL^{1/\nu})\\
\label{scal22}
N_2(\sigma,\epsilon,L)&  =   &  L^{\theta}  {\hat   N}_2  (B(\epsilon)
uL^{1/\nu}) \ .
\end{eqnarray}
As can be seen in Fig.~\ref{FIG10},  a very good collapse of the whole
set of curves  in the range $0.0005\le \epsilon  \le 0.03$ is obtained
with  the values  of $B$  indicated  in Table  \ref{Best} (this  scale
factor is arbitrarily set equal to $1$ for $\epsilon=0.008$).

A cross-check of the consistency  of the scaling collapses can be done
by fitting  the data in Fig.~\ref{FIG8}(b)  using Eq.~\ref{scalK} with
$\nu=1.2$  and $A=-0.1$.   The extrapolated  values of  $\sigma_c$ for
$L\rightarrow  \infty$ are  fully  compatible with  those reported  in
Table \ref{Best}.

\begin{figure}
\begin{center}
\epsfig{file=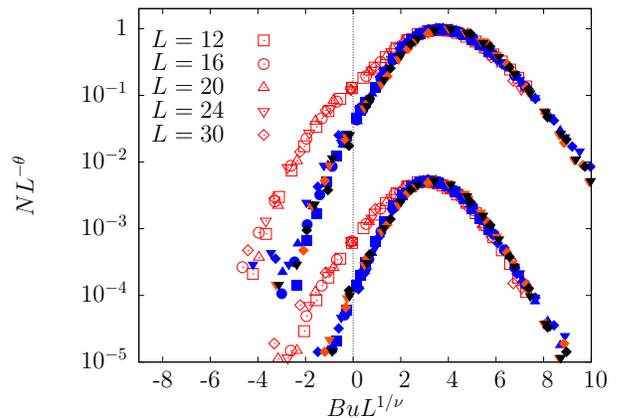, width=8cm,clip=}
\end{center}
\caption{\label{FIG11}  (Color  online)   Comparison  of  the  scaling
functions $\hat  N_1$ (above) and $\hat N_2$  (below) for $\epsilon=0$ (empty
symbols) and $\epsilon  >0$ (filled symbols) on a  linear-log scale. Different
symbols indicate the system  size and  different colors the  value of
$\epsilon$. $\hat N_2$ has been shifted two decades downwards for clarity.}
\end{figure}

Our data are  thus consistent with the fact  that the disorder-induced
critical point found for  $\epsilon=0$ transforms into a critical line
when $\epsilon > 0$ and the system is allowed to visit weakly unstable
states.  The whole  critical line appears to be  described by the same
exponents and by  the same scaling function for  $\epsilon>0$.  On the
other hand,  it seems that the  scaling function differs  from the one
corresponding  to  $\epsilon=0$, as  shown  in  Fig.~\ref{FIG11} on  a
linear-log scale  (in the figure, $B$  is chosen so as  to produce the
best collapse  on the right-hand  side of the peak;  for $\epsilon=0$,
this yields $B=1.35$ and $A=-0.1$ but the picture essentially does not
change with $A=-0.2$).

The fact  that the exponents $\nu$  and $\theta$ are the  same but the
scaling functions are different  for $\epsilon=0$ and $\epsilon>0$ may
be  surprising at  first  sight.  On  the  one hand,  this may  simply
indicate  that higher-order  terms  in the  scaling  variable $u$  are
needed   (let    us   recall   again   that    the   simplest   choice
$u=(\sigma-\sigma_c)/\sigma_c$ does not produce good scaling collapses
and  that it was  proposed already  in Ref.~\onlinecite{S1993}  to use
$u=(\sigma-\sigma_c)/\sigma$,  which amounts  to  keeping an  infinite
number    of     terms    in    an    expansion     in    powers    of
$(\sigma-\sigma_c)/\sigma_c$). It is  clear that simulations with much
larger  system   sizes  would  be   required  to  fully   settle  this
issue\cite{note6}.   On  the  other  hand,  there  may  be  indeed  an
essential  physical difference  in the  properties of  the percolating
clusters  at the  critical point.   This  may not  change the  fractal
dimension  (as this would  be reflected  in a  change of  the critical
exponents) but  only the  way the finite  size affects the  number and
size   of  these   clusters.   It   can  be   seen  for   instance  in
Fig.~\ref{FIG11}  that the  number of  1D- and  2D-spanning avalanches
diverges like $\sim L^{0.1}$ at  the critical point $u=0$ but that the
prefactor  of  this  divergence  dramatically  decreases  as  soon  as
$\epsilon$  becomes slightly  positive.  Finite-size  scaling functions
are known  to depend  sensitively on boundary  conditions and,  in the
present  case, they  may account  for  the fact  that the  percolating
clusters  are growing  in a  different environment  when $\epsilon>0$.
Recall that we  are dealing here with a  nonequilibrium phenomenon and
that, in  the original model\cite{PDS1999},  a finite fraction  of the
system  has already transformed  when the  critical point  is reached.
This is  reflected in the  value of the critical  magnetization $M_c$.
It is  not easy to estimate the  actual value of this  quantity in the
thermodynamic limit,  but preliminary calculations  suggest that $M_c$
significantly     decreases     when     $\epsilon>0$    (even     for
$\epsilon=0.0005$), showing that the  fraction of spins that flip when
driving   the  system   from   $H=-\infty$  to   the  critical   field
$H_c(\epsilon)$ becomes very small.

\section{Summary and conclusion}
\label{Conclusion}

We have studied the  zero-temperature random-field Ising model, with a
Gaussian  distribution   of  the  random  fields,   using  a  modified
single-spin-flip  dynamics that  allows the  system to  be  trapped in
weakly  unstable states  when driven  quasi-statically by  an external
field.  The  new dynamics,  however, does not  modify the  sequence of
states  that  are  visited  during  a  monotonous  field  history  and
preserves the  intermittent, avalanche-like character  of the response
to the driving  field.  The violation of the  standard local stability
condition is controlled by  a single parameter $\epsilon$ whose effect
is  somewhat similar  to that  of  a finite-driving  rate, moving  the
system away  from equilibrium and  resulting in a similar  increase of
the  width of  the  saturation  hysteresis loop,  as  measured by  the
coercive field.   Avalanches are modified but two  distinct regimes of
avalanches and two different loop shapes as a function of disorder are
still present.   As in the original  model\cite{S1993}, the transition
between  the  two  regimes  corresponds  to  a  critical  point  where
avalanches of all sizes are observed. The critical exponents that have
been extracted  from a finite-size  scaling analysis of the  number of
the  spanning  avalanches  appear  to be  independent  of  $\epsilon$,
suggesting that the condition of  strict metastability (as well as the
no-passing rule) may be irrelevant  for the critical behavior.  In our
opinion,  this significantly enlarges  the domain  of validity  of the
original model.

This work  has received financial support from  CICyT (Spain), project
MAT2007-61200,     CIRIT     (Catalonia),    project     2005SGR00969.
F.S.-P. acknowledges a post-graduate studies grant from the Fundaci\'o
``la Caixa''.


\begin{thebibliography}{10}
\bibitem{note2} For $\sigma=1.5$, there is a spanning avalanche in the
system  and the value  of $H_{coer}$  computed in  a small  system may
significantly differ from the actual value in the thermodynamic limit.
However,  for $L=24$, the  shift in  coercivity due  to the  change of
$\epsilon$ is much larger than the error due to finite size effects.
\bibitem{note5} Only for $\epsilon=0.0005$, the curve corresponding to
the smaller size $L=12$ could not be included in the collapse.
\bibitem{note6}  Note  that  there  are  typically none  or  very  few
spanning avalanches  along a half-loop.  Therefore, in  order to reach
good statistics for $N_1$ and $N_2$, it is not sufficient to analyze a
few realizations of very large size.
\end{thebibliography}
\end{document}